\documentclass[aps,prb,twocolumn,superscriptaddress,showpacs]{revtex4}

\usepackage{graphicx}

\begin{document}

\def\lufesi{Lu$_{2}$Fe$_{3}$Si$_{5}$}
\def\tc{$T_{c}$}
\def\mgb{MgB$_{2}$}
\def\lafe{LaFeAs(O,F)}
\def\spm{$s_{\pm}$}

\def\rfesi{$R_{2}$Fe$_{3}$Si$_{5}$}
\def\scfesi{Sc$_{2}$Fe$_{3}$Si$_{5}$}
\def\yfesi{Y$_{2}$Fe$_{3}$Si$_{5}$}
\def\tmfesi{Tm$_{2}$Fe$_{3}$Si$_{5}$}
\def\erfesi{Er$_{2}$Fe$_{3}$Si$_{5}$}


\title{Two-band superconductivity featuring different anisotropies in the ternary iron silicide {\lufesi} }


\author{Y.~Nakajima}
\affiliation{Department of Applied Physics, The University of Tokyo, Hongo, Bunkyo-ku, Tokyo 113-8656, Japan}
\affiliation{JST-TRIP, Hongo, Bunkyo-ku, Tokyo 113-8656, Japan}
\author{H.~Hidaka}
\author{T.~Nakagawa}
\affiliation{Department of Applied Physics, The University of Tokyo, Hongo, Bunkyo-ku, Tokyo 113-8656, Japan}
\author{T.~Tamegai}
\affiliation{Department of Applied Physics, The University of Tokyo, Hongo, Bunkyo-ku, Tokyo 113-8656, Japan}
\affiliation{JST-TRIP, Hongo, Bunkyo-ku, Tokyo 113-8656, Japan}
\author{T.~Nishizaki}
\author{T.~Sasaki}
\author{N.~Kobayashi}
\affiliation{Institute for Materials Research, Tohoku University, Katahira 2-1-1, Aoba-ku, Sendai 980-8577, Japan}



\date{\today}

\begin{abstract}
We report detailed studies of the upper critical field and low-temperature specific heat in the two-gap superconductor {\lufesi}. The anisotropy of the upper critical field suggests that the active band is quasi-one-dimensional. Low-temperature specific heat in magnetic fields reveals that the virtual $H_{c2}$ in the passive band is almost isotropic. These results strongly indicate that the two bands have two different anisotropies, similar to the typical two-gap superconductor MgB$_{2}$, and their interplay may be essential to the two-gap superconductivity in {\lufesi}. 
\end{abstract}


\maketitle



The recent discovery of iron-pnictide superconductors has attracted much interest because of  their high transition temperature\cite{kamih08}. One of the remarkable features in this system is the multiband superconductivity, which is in sharp contrast to the high-temperature cuprate superconductors\cite{singh08b}.  In iron-pnictide superconductors, fully gapped $s$-wave state with opposite signs in different Fermi surfaces ({\spm} wave) has been proposed theoretically\cite{mazin08,kurok08}. The paring symmetry could be induced by the interband scattering between electron and hole pockets due to antiferromagnetic spin fluctuations. In fact, several experimental results in favor of the {\spm} wave have been observed\cite{chris08,inoso10}. These facts provide great insight into the essential relation between the nature of the multiband structure and the superconducting paring mechanism, in contrast to the typical two-gap superconductor MgB$_{2}$, which has weak interband coupling. In this context, iron-pnictide superconductors have enriched our understanding of multiband superconductivity. Therefore, a detailed clarification of the multiband nature in superconductors is indispensable for elucidating not only the superconducting properties but also the pairing mechanism.

Ternary iron-silicide {\lufesi}, which with {\tc} $\sim$ 6 K has the highest transition temperature among iron-based superconductors other than the iron pnictides and iron chalcogenides, can be another candidate for a canonical multigap superconductor\cite{braun80}.  {\lufesi} crystallizes in the tetragonal {\scfesi}-type structure consisting of quasi-one-dimensional iron chains along the $c$ axis and quasi-two-dimensional iron squares parallel to the basal plane\cite{chabo84b}. Non-magnetic Fe $3d$ electrons in {\lufesi} should play a significant role in the superconductivity. The multigap superconductivity in {\lufesi} has been revealed by the detailed study of low-temperature specific heat\cite{nakaj08}. The experimental results reveal that the amplitudes of the larger gap $\Delta_{1}$ and the smaller one $\Delta_{2}$ are $2\Delta_{1}/k_{B}T_{c}=4.4$ and $2\Delta_{2}/k_{B}T_{c}=1.1$, respectively, and that each band contributes to the density of states almost equally. Specific heat,\cite{nakaj08} penetration depth,\cite{gordo08} and thermal conductivity studies\cite{machi11} suggest that the superconducting gaps are fully opened on the whole Fermi surface. However, interestingly, a rapid decrease in {\tc} is reported when a small amount of nonmagnetic impurities replace some of Lu or Si sites, \cite{hidak09,watan09,hidak10} which leads us to speculate that the sign of the superconducting gap functions is reversing on and/or between the Fermi surfaces such as the {\spm}-wave state. In order to reveal the peculiar superconducting properties in the multigap superconductor {\lufesi}, it is crucial to investigate the detailed nature of the multigap structure.

Study of anisotropy of the superconducting properties is important to clarify the multiband nature of the superconductivity.\cite{lyard04, okaza09} The anisotropy of the upper critical field can reflect directly the anisotropy of the active band, which has the larger gap. Low-temperature specific heat in magnetic fields is a useful probe for the low-lying excitation of quasi particles, which can reveal the details of the smaller gap in the passive band. Although thermal conductivity in magnetic fields, which also reflects low-lying quasiparticle excitations, has been measured,\cite{machi11} the anisotropy is not clarified. Therefore, both measurements can provide us with information on the nature of multiband superconductivity in {\lufesi}.

We report the upper critical field and low-temperature specific heat under magnetic fields in the multigap superconductor {\lufesi}. We find quasi-one-dimensional anisotropy in the active band with the larger superconducting gap and isotropy in the passive band with the smaller gap. These results imply that two different anisotropies may possibly play important roles in the two-gap superconductivity in {\lufesi}.


Single crystals of {\lufesi} were grown by the floating-zone technique using an image furnace and were annealed for four weeks at 1250$^{\circ}$C followed by five days at 800$^{\circ}$C. \cite{nakaj08} The resistivity measurements are performed by standard four-wire configuration. Specific heat in the magnetic field was measured by the relaxation method in a $^{3}$He refrigerator.

\begin{figure}[t]
\includegraphics[width=9cm]{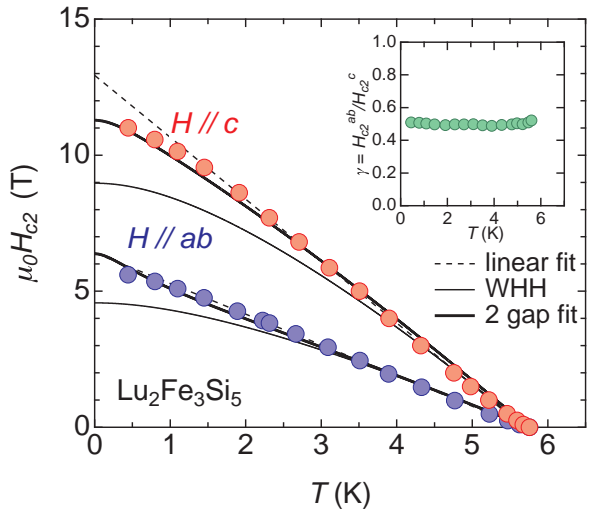}
\caption{(Color online) Temperature dependence of upper critical field for $H\parallel c$ and $H \parallel ab$ in {\lufesi}. Dashed lines represent the linear fit to the data. Thin lines are  the calculations based on the single band WHH theory. Thick lines show the calculations based on the two-gap model. \label{FIG1}}
\end{figure}


Figure 1 shows the temperature dependence of the upper critical field for $H\parallel ab$ and $H\parallel c$ in {\lufesi} obtained by the midpoint of resistive transition. In both directions, the upper critical fields increase almost linearly with decreasing temperature down to $\sim T_{c}/3$, which is strikingly different from conventional type II superconductors. In conventional superconductors, the upper critical field due to the orbital depairing is well described by the Werthamer-Helfand-Hohenberg (WHH) theory. \cite{werth66} In this theory, there is a simple universal relation between the zero-temperature value $H_{c2}(0)$ and the slope at $T_{c}$, $| dH_{c2}/dT |_{T=T_{c}}$, as follows: $H_{c2}(0)=\alpha T_{c} \left | \frac{dH_{c2}}{dT}\right |_{T=T_{c}}$, where $\alpha$ is 0.69 and 0.73 in dirty and clean limits, respectively. The thin lines in Fig. 1 are curves calculated based on the WHH theory in the dirty limit. Interestingly, the values of $H_{c2}(0)$ along the $c$ and $a $axies for {\lufesi} are about 11.5 T and 6 T, respectively, which correspond to $\sim 0.9T_{c}|dH_{c2}/dT|_{T=T_{c}}$ and are much larger than the values expected from WHH theory.

Since this upper critical field is so large, it may suggest spin-triplet superconductivity. However, such a possibility contradicts the Josephson effect results.\cite{noer85} We estimate the Pauli paramagnetic limiting field $H_{P}$. Simple calculation within the weak-coupling BCS theory with $\Delta = 1.76k_{B}T_{c}$ provides the well-known result $H_{P}^{BCS}$ (Tesla) = 1.84 $T_{c}$ (K). For simplicity, we estimate $H_{P}$ in analogy with a single-band superconductor.  In the strong electron-boson coupling case, $H_{P}$ can be modified as $H_{P} = (1+\lambda_{e\mathrm{-}b})H_{P}^{BCS}$ (Ref. \onlinecite{schos89}).  In {\lufesi} ($T_{c} = 5.8$ K), this limit is $10.7(1+\lambda_{e\mathrm{-}b})\sim 16.7$ T, where $\lambda_{e\mathrm{-}b}\sim 0.56$ is obtained from the specific heat results ($\Theta_{D} = 389$ K) using McMillan's formula with $\mu^{\ast} = 0.1$ (Ref. \onlinecite{mcmil68}).  The enhancement of $H_P$ can be also obtained by the following phenomenological rough estimation for a two-band superconductor. The larger gap with $\Delta_{1} = 2.2 k_{B}T_{c}$ \cite{nakaj08} determines condensation energy at low temperatures and high fields where the smaller one is almost suppressed. Therefore, $H_{P}$ should be enhanced and is $2.2 / 1.76\times H_{p}^{BCS} \sim$ 13.5 T. These estimations indicate that the upper critical field in {\lufesi} is determined by the orbital depairing.

We discuss how the orbital effect enhances the upper critical field as observed in the extended linear temperature dependence. To explain the excess of the WHH value, we attempted a preliminary analysis using a model describing the upper critical field for dirty two-gap superconductors \cite{nakaj09}. This model reproduces $H_{c2}$ as drawn by thick lines in Fig. 1. However, obtained coupling constants ($\lambda_{11} = 0.202, \lambda_{22} = 0.103, \lambda_{12}=0.059, \lambda_{21}= 0.053$, where $\lambda_{ii}$ and $\lambda_{ij}$ are intra- and inter-band coupling constants, respectively) do not provide the ratio of the gap amplitudes $\Delta_{1}/\Delta_{2}\sim 4$ obtained from specific heat measurements,\cite{nakaj08} penetration depth measurements,\cite{gordo08} and $\mu$SR results.\cite{biswa11}  Moreover, the preliminary results of dHvA effect reveal that the mean free path $l$ in the normal state is $\sim$ 1400 {\AA},\cite{teras11} which is much larger than coherence length $\xi_{ab}\sim$ 54 {\AA} and corresponds to the clean limit $l/\xi_{ab}>1$. These facts suggest that the two-gap model in the dirty limit may not be valid for this system. Another possibility for the enhanced $H_{c2}$ is due to Fermi surface topology. In clean $\beta$-pyrochlore KOs$_{2}$O$_{6}$, $H_{c2}(0)$ values larger than the WHH value and linear temperature dependence of $H_{c2}$, which are very similar to the behavior of $H_{c2}$ in {\lufesi}, have been reported \cite{shiba06}. The uncommon behavior of $H_{c2}$ in KOs$_{2}$O$_{6}$ is well described in a theoretical framework, where the orbital $H_{c2}$ is calculated based on the {\it ab initio} calculations of the details of Fermi surfaces responsible for superconductivity. The discrepancy of gap amplitudes obtained from the two-gap model for $H_{c2}$, the large mean free path, and the striking similarity of $H_{c2}$ between {\lufesi} and KOs$_{2}$O$_{6}$ lead us to conclude that the anomalous upper critical field in {\lufesi} could be due to the same origin as in KOs$_{2}$O$_{6}$.

The inset of Fig. 1 shows the anisotropy $\gamma= H_{c2}^{a}/H_{c2}^{c}$ as a function of $T$. $\gamma$ is almost independent of temperature. Its value is about 0.5, which reflects the weakly one-dimensional shape of the Fermi surface expected from the band structure calculation.\cite{nakaj08} We note that in the multiband superconductor with different anisotropies, such as MgB$_{2}$ and the iron-pnictide superconductors,\cite{lyard04,okaza09} $\gamma$ can be strongly temperature dependent due to the different anisotropies. The present results seem to imply that the bands with the larger and smaller gaps in {\lufesi} have similar anisotropies, but this is inconsistent with prediction of band calculation, which shows different anisotropies in different Fermi surfaces.  In multiband superconductors,  at $T=0$ K, $\gamma$ is mainly determined by the gap anisotropy of the active band responsible for superconductivity. In contrast,  according to GL theory  \cite{kogan02,miran03}, $\gamma$ at $T_{c}$ is given as $\gamma (T_{c})=<\Omega^{2}(v_{F}^{ab})^{2}>/<\Omega^{2}(v_{F}^{c})^{2}>$, where $<\cdots>$ denotes the average over the Fermi surface. $v_{F}^{ab}$ and $v_{F}^{c}$ are the Fermi velocities parallel and perpendicular to the $ab$ plane, respectively. $\Omega$ represents the gap anisotropy, which is also related to gap amplitude. Therefore, if inequality of Fermi velocities between bands is smaller than that for gap amplitude, $\gamma$ could be also determined by the anisotropy of the active band even near $T_{c}$. In fact, thermal conductivity measurements have reported such an inequality of Fermi velocities between bands.\cite{machi11}

\begin{figure}[t]
\includegraphics[width=8cm]{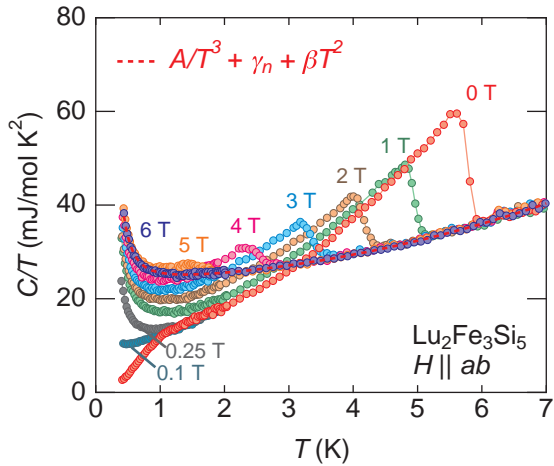}
\caption{(Color online) Specific heat $C/T$ at several fields with $H \parallel ab$ in {\lufesi} as a function of $T$. Dashed line is the fit to the data using $C/T = A/T^{3}+\gamma_{n}+\beta T^{2}$.\label{FIG2}}
\end{figure}


Figure 2 shows the specific heat $C/T$ for {\lufesi} as a function of $T$ at several fields along the $ab$ plane. It should be noted that at low temperatures in magnetic fields, upturn of $C/T$ is observed, which is possibly due to the Schottkey term, which originates from nuclear spins of Lu and/or impurities. In the normal state at $\mu_{0}H = $ 6  T along the $ab$ plane, the specific heat $C/T$ can be described by
\begin{equation}
	C/T= C_{Sch}(H,T)/T+\gamma_{n} +\beta T^{2},
\end{equation}
where the first term represents the Schottkey anomaly, the second term is the electronic part, and the last term corresponds to the phonon contribution. At temperatures higher than the energy scale of level splitting, the Schottkey anomaly term can be described as $C_{Sch}(H,T)/T =A(H)/T^{3}$.  From the fit to the data below 7 K,  we estimate $\gamma_{n}=24.1$ mJ/mol K$^{2}$ and $\beta = 0.331$ mJ/mol K$^{4}$, which is very similar to our previous results.\cite{nakaj08}

\begin{figure}[t]
\includegraphics[width=8cm]{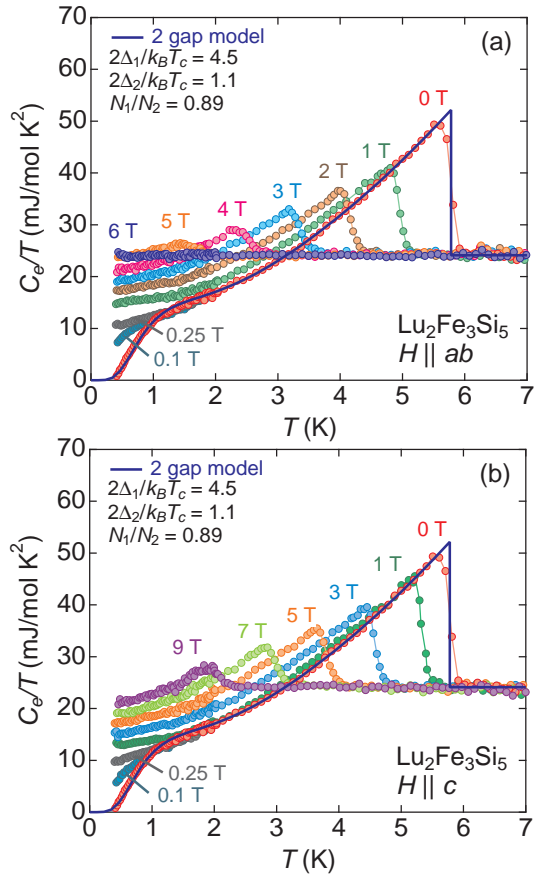}
\caption{(Color online) Temperature dependence of electronic specific heat $C_{e}/T$ for (a) $H\parallel ab$ and (b) $H \parallel c$ in {\lufesi}.  \label{FIG3}}
\end{figure}

Figure 3 shows the electronic specific heat $C_{e}/T$ obtained from the subtraction of Schottkey and phonon terms from $C/T$, as a function of $T$ at several magnetic fields.
We note that the fit to the data at 0 T  using the two-gap model, the so-called $\alpha$ model, gives almost the same values of $2\Delta_{1}/k_{B}T_{c}=4.5, 2\Delta_{2}/k_{B}T_{c}=1.1$, and $ N_{1}/N_{2}=0.89$ (Ref. \onlinecite{nakaj08}) as in the previous ones, although the sample here is a different piece obtained from the same batch as the previous study. Above 1.5 K, which corresponds to the energy scale of the smaller gap, $C_{e}/T$ increases almost linearly with increasing magnetic field. In contrast to this, below 1.5 K, a steep increase in $C_{e}/T$ at very low fields below 0.5 T is observed. This  behavior indicates that the smaller gap is suppressed by a very small field. We note that the striking field dependence at low fields and at low temperatures is also observed in MgB$_{2}$,\cite{lyard04} which confirms the presence of a distinct smaller superconducting gap in {\lufesi}.

\begin{figure}[t]
\includegraphics[width=8cm]{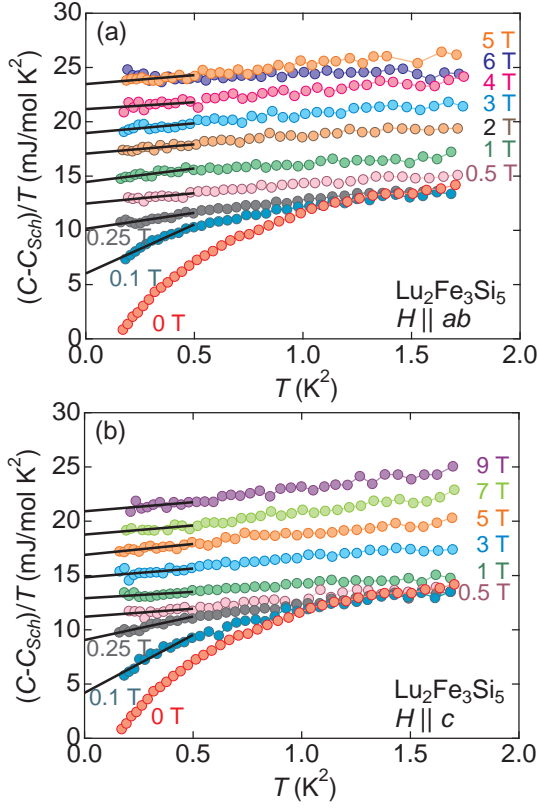}
\caption{(Color online) Specific heat corrected for the Schottkey contribution $(C-C_{Sch})/T$ as a function of $T^{2}$ for (a) $H\parallel ab$ and (b) $H \parallel c$ in {\lufesi}. The solid lines represent the linear fits. \label{FIG4}}
\end{figure}

\begin{figure}[t]
\includegraphics[width=8cm]{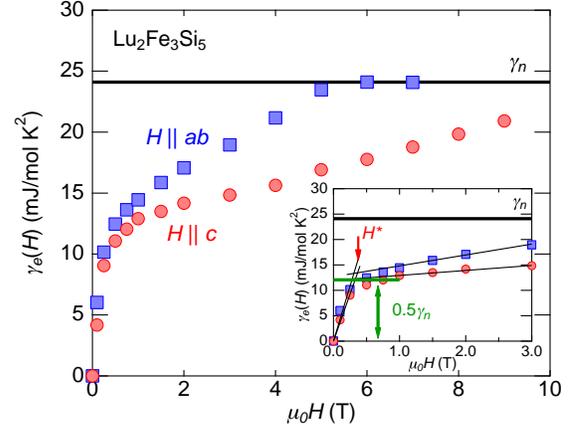}
\caption{(Color online) Field dependence of electronic specific heat coefficient $\gamma_{e}(H) = C_{e}/T$ for $H\parallel ab$ and $H \parallel c$ in {\lufesi}. The solid line represent the normal-state value $\gamma_{n}$. Inset: low-field part for $\gamma_{e}(H)$. \label{FIG5}}
\end{figure}

The magnetic field dependence of $C_{e}/T=\gamma_{e}(H)$ at low temperatures is instructive for the quasi-particle excitations. In order to extract the electronic specific heat in the low temperature limit, we plot the low-temperature $(C-C_{Sch})/T$ as a function of $T^{2}$ under several magnetic fields in Fig. 4. 
By extrapolating to $T=0$ K using a linear fit below $T^{2} = 0.5$ K$^{2}$, we can obtain the field dependence of $\gamma_{e}(H)$.

Figure 5 shows the field dependence of $\gamma_{e}(H) $ for $H\parallel ab$ and $H\parallel c$. $\gamma_{e}(H)$ shows a steep increase at very low fields and a kink at almost the same fields $\mu_{0}H^{\ast}\sim 0.33$ T in both field directions. Above the kink, $\gamma_{e}(H)$ increases almost linearly with magnetic fields to the normal-state value of $\gamma_{n}$. In fully gapped conventional superconductors, $\gamma_{e}(H)$ is proportional to $H$, which is due to the localized quasiparticle states in the vortex core. In contrast, for nodal superconductors, such as $d$-wave superconductors, $\gamma_{e}(H)$ is proportional to $\sqrt{H}$, which originates from extended quasiparticle states near nodal directions of the order parameter. The kink structure in $\gamma_{e}(H)$ at low fields observed in {\lufesi} is strikingly different from the field dependence of $\gamma_{e}(H)$ for both the conventional $s$-wave and the nodal superconductors but very similar to that for the typical two-gap superconductor {\mgb}.\cite{bouqu02} The steep increase indicates the suppression of the smaller gap by the magnetic field. The characteristic field $H^{\ast}$ corresponds to the virtual upper critical field of the smaller gap. Above $H^{\ast}$, the smaller gap is almost completely suppressed and the electrons in the passive band become normal. In fact, as shown in the inset of Fig. 5, $\gamma_{e}(H)$ at $H^{\ast}$ is $\sim 0.5\gamma_{n}$, which suggests that the contribution to the density of states from the passive band is half of the total density of states. This is consistent with the independent results of $N_{1}/N_{2}\sim 1$ obtained from the temperature dependence of electronic specific heat at 0 T and the penetration depth measurements \cite{gordo08}.  It should be emphasized that the isotropic $H^{\ast}$ along the $c$ axis and $ab$ plane indicates that the passive band is isotropic, which suggests that the two bands have different anisotropies.

The {\it virtual} $H_{c2}$ and upper critical field can provide us with a simple estimation of the ratio of the magnitude of two gaps.  The {\it virtual} $H_{c2}$ of the smaller gap can be written as $H^{\ast}\sim \Phi_{0}/2\pi \xi_{2}^2$ and the upper critical field along the $c$ axis can be written as $H^{c}_{c2}\sim \Phi_{0}/2\pi (\xi_{1}^{ab})^{2}$. Here, $\xi_{1}^{ab}=\hbar v_{F1}^{ab}/\pi\Delta_{1}$ and $\xi_{2}=\hbar v_{F2}^{ab}/\pi\Delta_{2}$ are the coherence lengths along the $ab$ plane for the active band and for the passive band, respectively, and $v_{Fi}^{ab} (i = 1,2)$ is the Fermi velocity along the $ab$ plane for each band. We note that the thermal conductivity measurements,\cite{machi11} which reflect the lighter carriers, suggest the inequality of carrier masses with two bands and the ratio of isotropic Fermi velocities $v_{F1}/v_{F2}\sim 0.8$. We get the ratio of gap amplitude $\Delta_{1}/\Delta_{2}=\sqrt{H_{c2}^{c}/H^{\ast}} (v_{F1}^{ab}/v_{F2}^{ab})\sim 4.7$ (Ref. \onlinecite{machib}), which is also similar to the independent results obtained from the temperature dependence of electronic specific heat at 0 T and the penetration depth measurements.


Summarizing the results, we find a quasi-one-dimensional active band and a three-dimensional isotropic passive band in {\lufesi}. According to the band calculation \cite{nakaj08}, Fermi surfaces are composed of two hole bands with quasi-one-dimensional parts and an electron band with a three-dimensional shape. The contribution of the two hole bands to the total density of states is 58.4 \% while that of the electron band is 41.6 \%. These facts suggest that the three-dimensional electron band is possibly the passive band and the quasi-one-dimensional hole bands are the active band. We speculate that the two-different anisotropies in these bands possibly play an important role in the two-gap superconductivity in {\lufesi} because orthogonality between these bands, which could originate from the difference in dimensionality, leads to moderate interband coupling as in the case of {\mgb}.

One may speculate that the irons with different site symmetries (namely, 4$d$ and 8$h$) play an important role in giving rise to multiband superconductivity. The irons on the 8$h$ sites form an octahedron with six Si atoms and the irons on the 4$d$ sites form a tetrahedron with four Si atoms. According to the band calculation,\cite{harim11} the contribution of the 4$d$ and 8$h$ Fe atoms to the partial density of states for all the bands is almost the same. Therefore, it is important to investigate the orbital-dependent contribution to each band in more detail.


In summary, we measure the upper critical field and low-temperature specific heat in the two-gap superconductor {\lufesi}. Anisotropy of the upper critical field suggests that the active band is quasi-one-dimensional. The isotropy of the {\it virtual} upper critical field $H^{\ast}$ obtained by field dependence of low-temperature specific heat indicates that the passive band is isotropic. The difference of the dimensionality between the passive and active bands may play an important role in giving rise to the multi band superconductivity.

\begin{acknowledgments}
We thank H. Harima, T. Terashima, T. Shibauchi, K. Izawa, T. Watanabe, and Y. Machida for useful discussions and N. Shirakawa for technical assistance in thermometer calibration under high magnetic fields. This work was partly supported by a Grant-in-Aid for Scientific Research from MEXT, Japan. 
\end{acknowledgments}


\begin{thebibliography}{29}
\bibitem[{mac()}]{machib} We assume $v_{F1}^{ab}/v_{F2}^{ab} \sim v_{F1}/v_{F2}$ because an averaged Fermi velocity for each band have been obtained from thermal conductivity measurements assuming an isotropic $v_F$ in ref. 11.
\end{thebibliography}
\end{document}